\documentclass[12pt,preprint]{aastex}

\shorttitle{MHD wave heating of cool cores}
\shortauthors{Fujita et al.}

\begin{document}

\title{The Origin of Ripples in Cool Cores of Galaxy Clusters: Heating
by MHD Waves?}

\author{Yutaka Fujita\altaffilmark{1}, Takeru K. Suzuki\altaffilmark{2}, 
Takahiro Kudoh\altaffilmark{3},
and Takaaki Yokoyama\altaffilmark{4}}

\altaffiltext{1}{Department of Earth and Space Science, Graduate School
of Science, Osaka University, 1-1 Machikaneyama-cho, Toyonaka, Osaka
560-0043, Japan; fujita@vega.ess.sci.osaka-u.ac.jp}
\altaffiltext{2}{Graduate School of Arts and Sciences, 
University of Tokyo, Komaba, Meguro, Tokyo 153-8902, Japan; 
stakeru@ea.c.u-tokyo.ac.jp}
\altaffiltext{3}{Division of Theoretical Astronomy, National 
Astronomical Observatory of Japan, Mitaka, Tokyo 181-8588, Japan;
kudoh@th.nao.ac.jp}
\altaffiltext{4}{Department of Earth and Planetary Science, 
University of Tokyo, Hongo, Bunkyo, Tokyo, 113-0033, Japan; 
yokoyama.t@eps.s.u-tokyo.ac.jp}

\begin{abstract}
 We consider MHD waves as a heating source of cool cores of galaxy
 clusters. In particular, we focus on transverse waves (Alfv\'en waves),
 because they can propagate a longer distance than longitudinal waves
 (sound waves). Using MHD simulations, we found that the transverse
 waves can stably heat a cool core if the wave period is large enough
 ($\gtrsim 10^8$~yr). Moreover, the longitudinal waves that are created
 as a by-product of the nonlinear evolution of the transverse waves
 could be observed as the 'ripples' found in cool cores.
\end{abstract}


\section{Introduction}

X-ray observations have shown that the radiative cooling time of the ICM
(intracluster medium) in the central regions of galaxy clusters (cool
cores) is generally smaller than the Hubble time. If there were no
heating sources, the ICM flows subsonically toward the cluster center
with a mass deposition rate of $\sim 100$--$1000\: M_\sun\:\rm yr^{-1}$
\citep{fab94}. This flow had been called a 'cooling flow'. However,
recent X-ray observations have shown that the actual cooling rate of the
ICM is much smaller. The Japanese {\it ASCA} (Advanced Satellite for
Cosmology and Astrophysics) team indicated that metal emission lines
from the low temperature cooling gas were much weaker than had been
predicted by the classical cooling flow model \citep{ike97,mak01}. This
has been confirmed by {\it XMM-Newton} \citep{pet01,kaa01,tam01}; the
actual mass deposition rate is at least 5 or 10 times less than that
previously assumed.

The lack of the cooling flows indicates that there are heating sources
in the central regions of clusters that prevent the development of
cooling flows. Most popular candidate of the heating source is AGNs at
the centers of clusters. However, although the energy released by the
AGNs is sufficient to cancel radiative cooling of the ICM in cool cores,
the mechanism of energy transfer from the AGNs to the surrounding ICM is
not clear. One possible mechanism is the dissipation of sound waves and
weak shocks that have evolved from sound waves
\citep{fab03,rus04,fab05}. Those waves could be generated by the
activities of the central AGNs of clusters. In fact, sound waves and
weak shocks are observed in the central regions of the Perseus and the
Virgo Clusters \citep{fab03,for05}.

However, it has been shown that sound waves (longitudinal waves) and
weak shocks that have evolved from the sound waves cannot stably heat a
cool core, because they should dissipate before they propagate in the
whole core.  In other words, their dissipation length is too small.
Even if their net energy is large enough to balance radiative cooling,
the temperature profile of a cluster soon becomes irregular, which is
not consistent with observations \citep{fuj05,mat06}. On the other hand,
it seems to be curious that structures that look like sound waves
('ripples') have been observed far away ($r\sim 50$~kpc) from the
central AGN in the Perseus cluster \citep{fab06}. Since the dissipation
length of the waves is small, the density fluctuations or the ripples
should not be observed there if they were created at the cluster center.

So far, only longitudinal waves have been considered as heating sources
in clusters. However, since the ICM is magnetized \citep*[$\sim
5$--$10\: \mu\rm G$;][]{cla01a}, there should be magnetohydrodynamic
(MHD) waves including tranverse waves. It has been known that transverse
waves (Alfv\'en waves) can propagate a longer distance than longitudinal
waves \citep[e.g][]{suz02}. Thus, the transverse waves could heat the
whole core. In this letter, we consider the heating by tranverse waves
and longitudinal waves that are a by-product of nonlinear evolution of
the transverse waves. The longitudinal waves could be observed as the
ripples. We do not restrict the wave generators only to AGNs, because a
long wave period is required to reproduce the ripples as shown below. It
is to be noted that \citet{mak01} proposed a heating model of MHD
effects, but they considered magnetic reconnection instead of waves.

\section{Models}

The models presented here are based on heating models of the solar
corona \citep{kud99,suz05}. We study the propagation of waves along a
magnetic field line extending from the cluster center in the radial
direction. We assume axisymmetry about this field line.  In the vicinity
of the line, we consider a local orthogonal curvilinear coordinate
system defined by $r$, the distance from the cluster center, by the
azimuthal angle $\theta$ measured about the axis of symmetry, and by
$\xi$ measured in the direction perpendicular to both $r$ and $\theta$
vectors \citep*{hol82}. In the following, the indices, $r$, $\theta$,
and $\xi$, indicate the vector components for each direction.

We assume ideal MHD (inviscid, perfect conductor) in a whole region:
\begin{equation}
   {\partial \over \partial t} (\rho S)
    + {\partial \over \partial r} (\rho v_r S)
    = 0\:,
\end{equation}
\begin{equation}
   {\partial \over \partial t}(\rho v_r S)
    + {\partial \over \partial r} \left[
      \left(\rho v_r^2 +p + {B_\theta^2\over{8\pi}}\right)S
    \right]
    = \rho \left[
    g_r + \left(v_\theta^2 - {B_\theta^2 \over 4 \pi \rho}\right) 
{1 \over R}{dR \over dr}
    \right] S
    + \left( p +  {B_\theta^2\over{8\pi}}\right) {dS\over dr}\:,
\end{equation}
\begin{equation}
   {\partial \over \partial t}(\rho v_\theta R S)
    + {\partial \over \partial r} \left[
      \left(\rho v_r v_\theta - {B_r B_\theta\over{4\pi}}\right)R S
    \right]
    = 0\:,
\end{equation}
\begin{equation}
   {\partial \over \partial t} \left({B_\theta S \over R}\right)
    - {\partial \over \partial r} \left({E_\xi S \over R}\right)
    = 0\:,
\end{equation}
\begin{equation}
  {\partial \over \partial t} \left[ \left(
       {p\over \gamma -1} + {1\over 2}\rho v^2
        + {B_\theta^2\over 8\pi}
     \right) S \right]
 + {\partial \over \partial r} \left[
       \left(\left({\gamma\over \gamma -1} p
        + {1\over 2}\rho v^2\right) v_r
        -{B_\theta E_\xi\over 4\pi}
    \right) S \right]
   = (\rho g_r v_r-L) S\:,
\end{equation}
\begin{equation}
    E_\xi = -v_rB_\theta+v_\theta B_r\:,
\end{equation}
\begin{equation}
    v^2 = v_r^2+v_\theta^2\:,
\end{equation}
where $\rho$ is the ICM density, $S$ is the cross section of the
magnetic flux tube, $v$ is the velocity, $p$ is the pressure, $B$ is the
magnetic field, $R$ is the distance from the axis of the symmetry, $E$
is the electric field, $\gamma (=5/3)$ is the adiabatic index, $g$ is
the gravitational acceleration, and $L$ is the cooling rate. We assumed
that $\partial/\partial\theta = 0$, $v_\xi=0$, and $B_\xi$ = 0 in order
to reduce the problem to one-dimension \citep{hol82}. For the sake of
simplicity, we assume that the cluster is spherically symmetric, which
means $S\propto r^{-2}$. The initial magnetic field is given by
$B_r\propto r^{-2}$ and $B_\theta =0$. We set the inner and outer
boundaries at $r_{\rm in}=5$~kpc and $r_{\rm out}=84$~Mpc,
respectively. Because of the large outer boundary, we do not have to
consider the reflection of waves there. We adopted the relatively large
inner radius to avoid the divergence of $B_r$ at the cluster center. The
beta parameter of the plasma is set to be one at $r=r_{\rm in}$. Thus,
for $r>r_{\rm in}$ it increases rapidly as $r$ increases, which means
that the Alfv\'en velocity $v_A$ is smaller than the sound velocity
$c_s$. At $r=20$~kpc, we obtain $B_r=6\:\mu\rm\: G$.

Transverse waves are injected around the inner boundary in a form of
acceleration in the $\theta$ direction. The wave acceleration is given
by
\begin{equation}
 A_w(t,r) = a g_{\rm in}\exp\left[-\frac{1}{2}
\left(\frac{r-r_{\rm in}}{r_{\rm in}}\right)^2\right]
\frac{\pi}{2}\sin\left(\frac{2\pi t}{P}\right)\:,
\label{eq:acc}
\end{equation}
where $g_{\rm in}$ is the gravitational acceleration from the cluster
potential at $r=r_{\rm in}$, and $a$ and $P$ are the parameters.
Longitudinal waves are not injected for simplicity. It is to be noted
that even if we give longitudinal waves with the same acceleration as
that of the transverse waves, the resultant gas velocity of the
longitudinal waves ($v_r$) is much smaller than that of the transverse
waves ($v_\theta$). This is because of the pressure gradient of the
cluster, which prevents the ICM from being lifted in the $r$-direction.

We adopt a cooling function based on the detailed calculations by
\citet{sut93},
\begin{equation}
\label{eq:cool}
 L=n_e^2 \Lambda = [C_1 (k_B T)^\alpha + C_2 (k_B T)^\beta 
+ C_3]n_i n_e\:,
\end{equation}
where $n_e$ and $n_i$ are the electron and the ion number densities,
respectively, and the units for $k_B T$ are keV. For an average
metallicity $Z=0.3\: Z_\sun$, the constants in equation (\ref{eq:cool})
are $\alpha=-1.7$, $\beta=0.5$, $C_1=8.6\times 10^{-3}$, $C_2=5.8\times
10^{-2}$, and $C_3=6.4\times 10^{-2}$, and we can approximate $n_i
n_e=0.70\:(\rho/m_H)^2$, where $m_H$ is the hydrogen mass. The units of
$\Lambda$ are $10^{-22}\:\rm ergs\: cm^3$ \citep{rus02}.

The model cluster we considered is similar to that adopted in \S~3.2 of
\citet*{fuj04a}. The mass density profile of the cluster is given by the
NFW profile \citep*{nav97}. The viral mass is $1.2\times 10^{15}\:
M_\sun$, the concentration parameter is 4.7, and the characteristic
radius is 460~kpc. For this cluster, $g_{\rm in}=1300\rm\: km\:
s^{-1}(10^8 \: yr)^{-1}$, which is a factor of two larger than the
acceleration observed at the cold front in A1795
\citep*{mar01}. Initially, the cluster is isothermal and the temperature
is 7~keV. The initial electron density at the cluster center is
$n_{e0}=0.017\rm\: cm^{-3}$. All of the calculations presented in this
letter use CANS (Coordinated Astronomical Numerical
Software)\footnote{http://www-space.eps.s.u-tokyo.ac.jp/\%7Eyokoyama/etc/cans/index-e.html}. The
number of grid points is 16392. The grid size is 8.3~pc for $r<120$~kpc
and it increases logarithmically for $r>120$~kpc.

\section{Results}
\label{sec:results}

\begin{figure}
\plotone{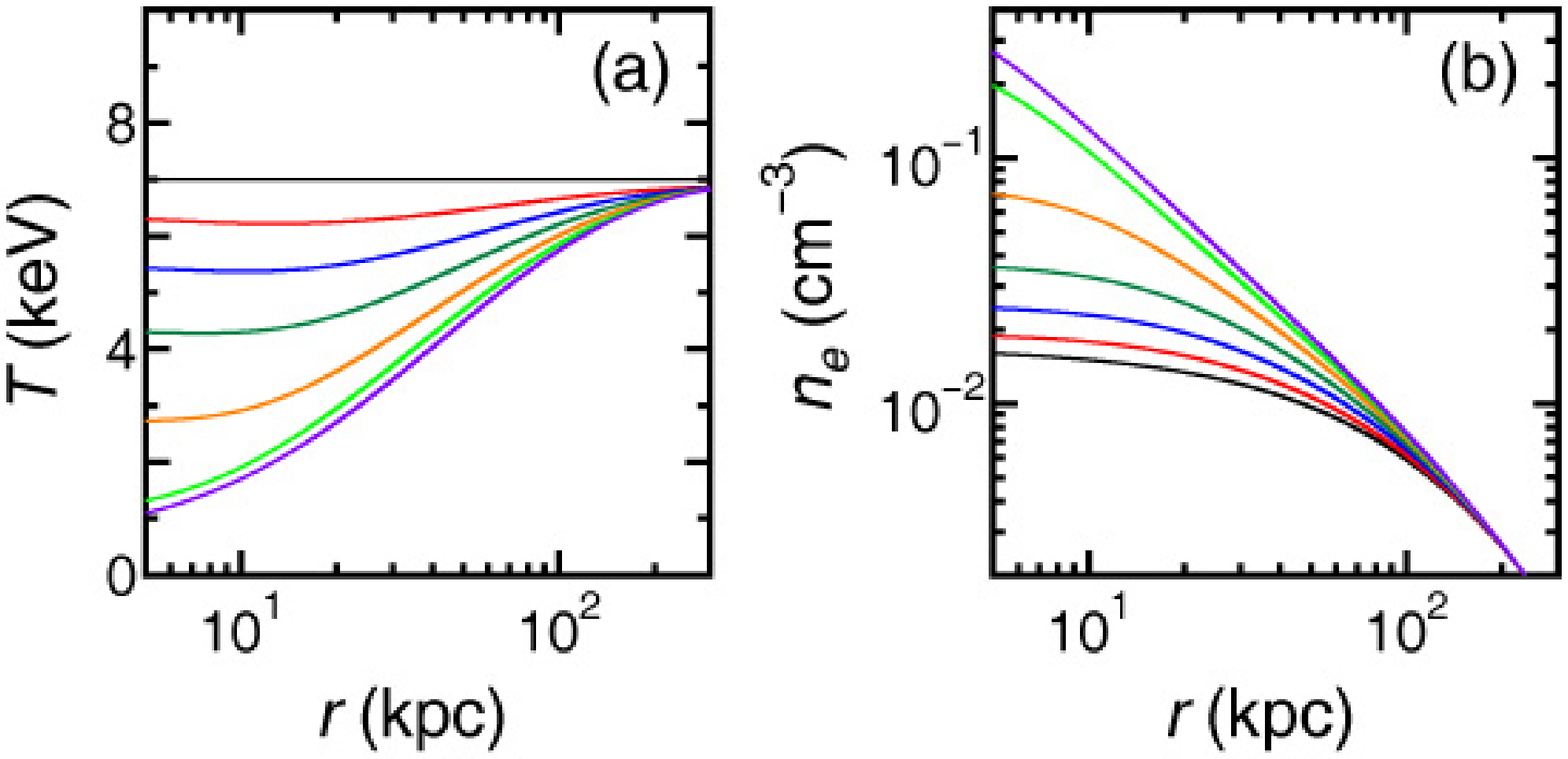} \caption{Time sequence of temperature and electron
number density for $a=0$. They are shown every 0.8~Gyr \label{fig:cf}}
\end{figure}

First, we investigate the case where waves are not injected for
comparison (a genuine cooling flow). Fig.~\ref{fig:cf} shows the
evolution of temperature and electron number density as functions of
distance from the cluster center. The interval of the lines is
0.8~Gyr. Because of radiative cooling, the temperature decreases and the
density increases monotonically at the cluster center.

Next, we consider the heating by Alfv\'en waves by injecting transverse
waves at the inner boundary. We take $a=1$ and $P=0.5$~Gyr.  For this
wave period $P$, the wave length of the sound waves is $c_s P\sim
700$~kpc. On the other hand, the wave length of the Alfv\'en waves is
smaller and is $v_A P \sim 10$~kpc at $r\sim 20$--50~kpc, which is
comparable to the observed wave length of ripples \citep{fab06}. This
wave period may be too large to be attributed to the activities of the
central AGN; it could be attributed to gas motion around the central
galaxy induced by minor cluster mergers such as 'sloshing' \citep{mar01}
or 'tsunami' \citep*{fuj04c}. The wave injection in a form of
acceleration (eq.~[\ref{eq:acc}]) would be preferable to model the waves
of this kind, because the acceleration is independent of the gas flow
toward the cluster center and the AGN. The transverse waves would be
generated on the side of the sloshing central galaxy, but not in front
and in the rear.

\begin{figure}
\plotone{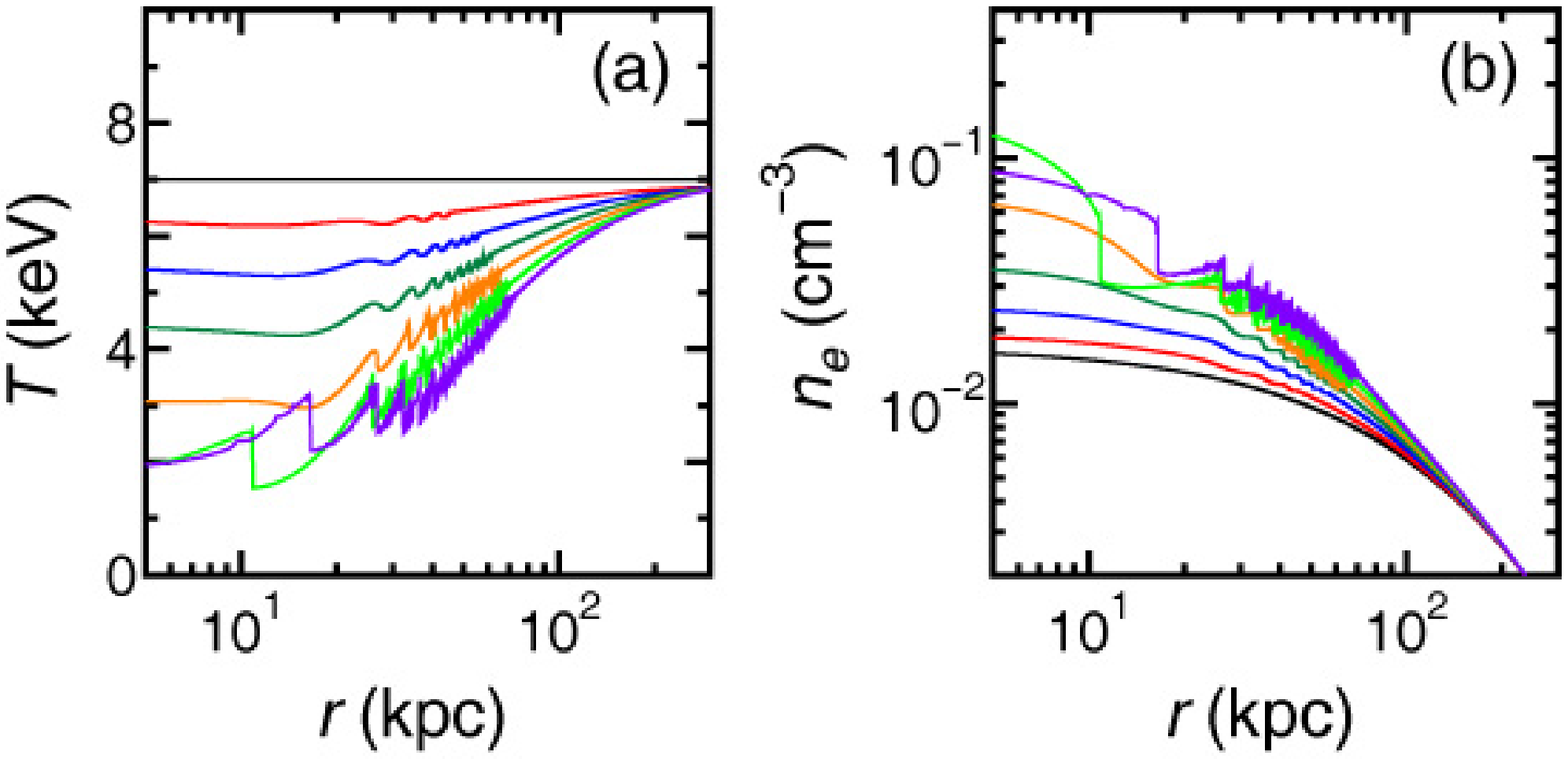} \caption{Same as Fig.~\ref{fig:cf}, but for $a=1$ and
 $P=0.5$~Gyr. \label{fig:tero_a1}}
\end{figure}

Fig.~\ref{fig:tero_a1} shows the evolution of the temperature and
density profiles when transverse waves are considered. We found that
while the temperature at the cluster center decreases for $t\lesssim
4$~Gyr, it is stable and almost constant for $t\gtrsim 4$~Gyr until we
stopped the calculation at $t=6$~Gyr. Since we inject waves in a form of
acceleration irrelevant of the gas mass of the core, the wave energy
increases as the gas density in the core and thus the gas mass of the
core increases. This is because for a given acceleration, the heavier
the gas core is, the larger kinetic energy is given to the gas
core. This is one reason that the temperature stops decreasing. The wave
energy flux per unit area is given by $F_A=\rho v_\theta^2(v_A+v_r)+v_r
B_\theta^2/(8\pi)$. At $r=15$~kpc (outside the wave injection region), the
energy flux is $4\pi r^2 F_A \sim 10^{42}\rm\: erg\: s^{-1}$ at $t\sim
0$, and it increases to $\sim 10^{44}\rm\: erg\: s^{-1}$ at $t\sim
6$~Gyr. Numerical simulations showed that the energy can be supplied by
minor cluster mergers \citep{tit05}.

Fig.~\ref{fig:vxvy_a1} shows the longitudinal and transverse velocity
profiles at $t=1.6$ and 3.2~Gyr. The transverse waves have reached
$r=57$~kpc at $t=1.6$~Gyr and $r=66$~kpc at $t=3.2$~Gyr. In
Fig.~\ref{fig:vxvy_a1}b, sharp jumps are seen; these jumps are formed
through the steeping of Alfv\'en waves and are called 'Switch-on shocks'
\cite[e.g.][]{hol82a,suz04}. At these shocks, while entropy is
generated, longitudinal waves are also generated through nonlinear
coupling \cite[e.g.][]{hol92, kud99, suz02, suz05}. The longitudinal
waves (sound waves) soon steepen and turn into weak shocks ({\sf
N}-waves; Fig.~\ref{fig:vxvy_a1}a), which also contribute to the heating
of the ICM. Note that $v_r$ is much smaller than $v_\theta$.  Contrary
to the transverse waves, the longitudinal waves can be observed in
temperature and density profiles. The temperature and density
fluctuations seen in Fig.~\ref{fig:tero_a1} reflect the generation of
the longitudinal waves. In particular, the density fluctuations could be
observed as ripples. Since these longitudinal waves are generated away
from the cluster center, they can be observed even at $r>50$~kpc
(Fig.~\ref{fig:tero_a1}b).

We also studied the mass inflow rate at the inner boundary. In the
cooling flow case (Fig.~\ref{fig:cf}), the inflow rate at $t=4.8$~Gyr is
$\dot{M}=220\: M_\sun\:\rm yr^{-1}$, and it is $310\: M_\sun\:\rm
yr^{-1}$ at $t=6$~Gyr. On the other hand, waves can decrease the inflow
rate significantly. In the case of Fig.~\ref{fig:tero_a1}, it is
$\dot{M}=60\: M_\sun\:\rm yr^{-1}$ at $t\sim 4.8$~Gyr, which means that
the waves effectively heat the surrounding ICM, although they cannot
completely shut down a cooling flow. The inflow rate gradually
increases, but even at $t\sim 6$~Gyr, it is still $\dot{M}=80\:
M_\sun\:\rm yr^{-1}$.

\begin{figure}
\plotone{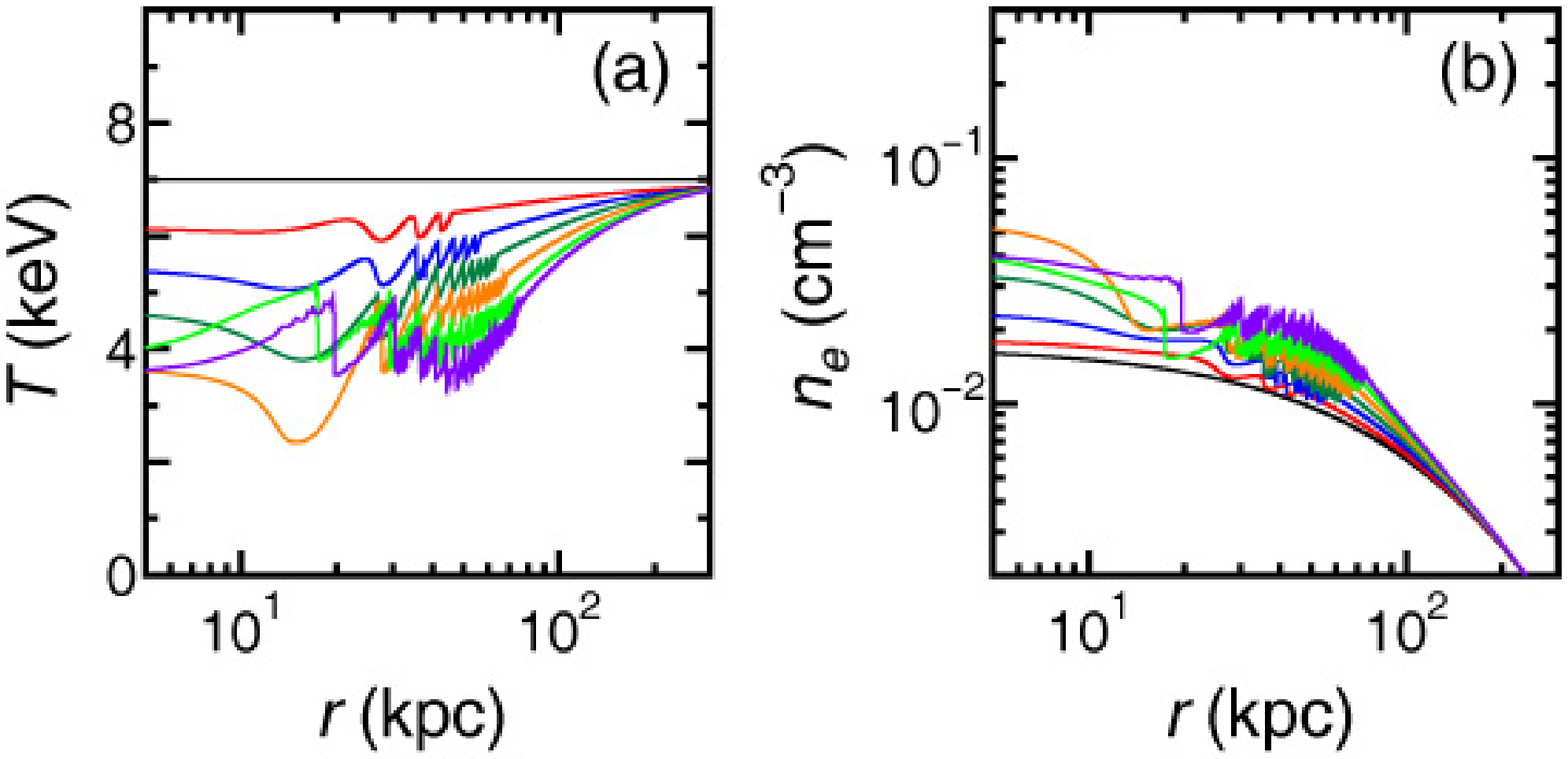} \caption{Same as Fig.~\ref{fig:cf}, but for $a=2$ and
 $P=0.5$~Gyr. \label{fig:tero_a2}}
\end{figure}

We found that the heating by the Alfv\'en waves is fairly stable even if
we change the acceleration parameter $a$, because as we mentioned above,
high density coming from radiative cooling results in a high heating
rate, and vice versa. Fig.~\ref{fig:tero_a2} shows the results when
$a=2$ and $P=0.5$~Gyr. Although the central temperature at which it
stops decreasing ($\sim 4$~keV; Fig.~\ref{fig:tero_a2}a) is higher than
that in the case of $a=1$ ($\sim 2$~keV; Fig.~\ref{fig:tero_a1}a), the
heating balances cooling after $t\sim 4$~Gyr. However, if $a$ is much
smaller than one, the waves cannot prevent the development of a cooling
flow. In summary, $a$ must be tuned to be $\sim 1$--2 in order to
reproduce observed temperature profiles.

Moreover, we have investigated the heating by Alfv\'en waves with a
relatively small wave period ($P\sim 10^7$~yr), which would be
appropriate for waves generated by the activities of the central AGN.
We found that the heating by these waves is not stable. The reason is
similar to the heating by sound waves alone, that is, the waves can heat
only the innermost region of the cluster.

\section{Summary and Discussion}

We have shown that Alfv\'en waves (transverse waves) with a long wave
period can effectively heat cool cores of galaxy clusters. Since the
dissipation rate of the waves is small, the waves can reach the outer
core and then heat the whole core before they dissipate. During the
propagation, sound waves (longitudinal waves) are generated through
nonlinear coupling, and they would be observed as ripples.  The long
dissipation length indicates the superiority of the heating by the
Alfv\'en waves over that by sound waves alone. In real clusters,
however, it would be possible that sound waves created by AGN activities
heat the inner cores, while Alfv\'en waves induced by cluster mergers
heat the outer cores. In this case, the wave acceleration ($A_w$)
required to heat the cores will be much smaller than that we assumed.
We assumed that the Alfv\'en waves are generated close to the cluster
center ($r\sim 5$--10~kpc). \citet{fuj04c} showed that complex gas
motion should be created even at the cluster center by minor cluster
mergers. Thus, our assumption may be justified. However, if the waves
are generated at some distance from the center, another heating source
is required to heat the center.

\begin{figure}
\plotone{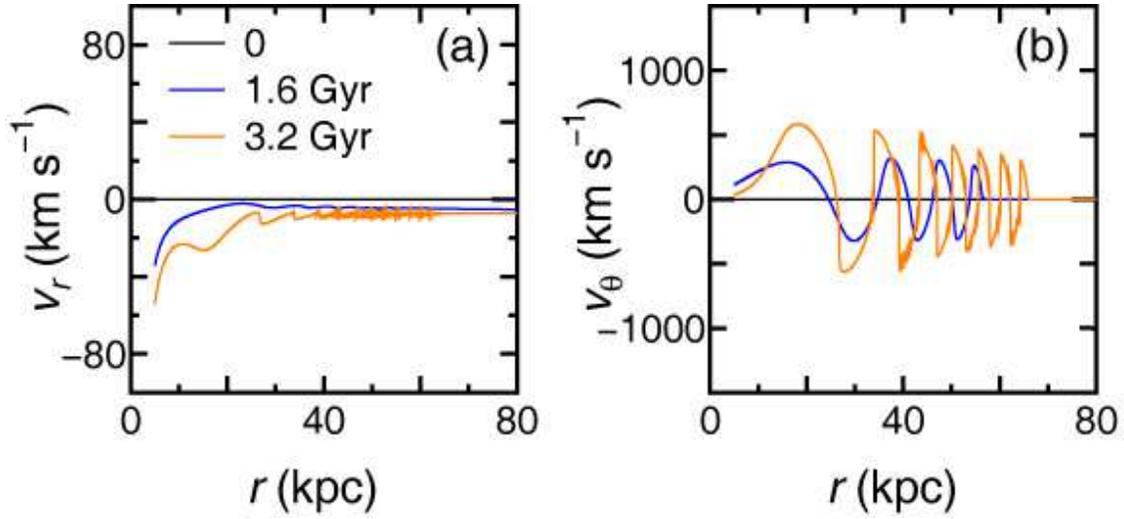} \caption{The longitudinal ($v_r$) and transverse
 ($v_\theta$) velocity profiles at $t=1.6$ and 3.2~Gyr. Note that the
 scales of the vertical axes are different. \label{fig:vxvy_a1}}
\end{figure}

In this letter, we considered only spherically symmetric magnetic
fields. If they are not spherically symmetric, the heating should
spatially be biased. In the region with stronger magnetic fields,
Alfv\'en waves propagate faster and the heating is more effective. This
could result in multiphase gas observed in the Perseus cluster
\citep{fab06}. If the longevity of optical filaments observed in the
Perseus cluster \citep*{con01} is smaller than the wave period, the
filaments would indicate the direction of the wave oscillation or the
motion of the central galaxy relative to the surrounding ICM. That is,
the waves would be propagating in the direction perpendicular to the
optical filaments. The model presented here predicts relatively large
transverse velocity fluctuations (Fig.~\ref{fig:vxvy_a1}b), which could
be detected in near-future space X-ray missions such as {\it NeXT}.

\acknowledgments

Y.~F. and T.~K.~S.\ were supported in part by a Grant-in-Aid from the
Ministry of Education, Culture, Sports, Science, and Technology of Japan
(Y.~F.: 17740162; T.~K.~S. 1884009).

\clearpage

\end{document}